\documentclass[lettersize,journal]{IEEEtran}
\usepackage{subfiles}
\usepackage{authblk}
\usepackage{graphicx}
\usepackage{amsmath,amssymb,amsfonts}
\usepackage{algorithmic}
\usepackage{textcomp}
\usepackage{xcolor}
\usepackage[normalem]{ulem}

\usepackage{soul}
\usepackage{url}
\usepackage{siunitx}
\usepackage{geometry}
\usepackage{subcaption}
\usepackage{svg}
\usepackage{multirow}
\usepackage{fancyhdr}
\usepackage{lastpage}
\usepackage{lineno}
\usepackage[labelfont=bf]{caption}
\usepackage[capitalise]{cleveref}
\usepackage[labelformat=simple]{subcaption}

\usepackage[sorting = none, backend = biber, style=numeric-comp]{biblatex}
\title{Node-reduction through Joint Optimization of Input and Readout Layers in Photonic Reservoir Equalization}
\author[1,2]{Ruben Van Assche}
\author[1,2]{Sarah Masaad}
\author[1,2]{Peter Bienstman}

\affil[1]{Photonics Research Group, INTEC, Ghent University - imec, 9052 Ghent, Belgium.}
\affil[2]{imec, Kapeldreef 75, 3001 Leuven, Belgium.}
\begin{document}

\maketitle
\begin{abstract}
Photonic reservoir computing is a machine learning paradigm in which a recurrent neural network remains fixed while only the output weights are trained. This makes it a well-suited approach for high-speed signal equalisation in optical communication systems, offering a trainable, low-power, and low-complexity solution. However, achieving strong performance typically requires relatively large network sizes, as learning is confined to the output layer. To address this, we investigate the role of trainable input mappings alongside conventional output weight optimisation. Across a range of short- and mid-reach IM/DD transmission scenarios, reaching up to 200 km for a 28 GBd NRZ signal, improvements of over two orders of magnitude in BER are achieved. This enables halving the network size while maintaining comparable performance. Furthermore, we show that this approach effectively extends the memory of the reservoir, resulting in over three orders of magnitude improvement in memory-intensive tasks. These results also show that starting at 16 nodes a performance of at least one to two magnitudes better than both a complexity matched FFE and a Volterra filter of second order are reached.
\end{abstract}

\section{Introduction}
Intensity-modulation/direct-detection (IM/DD) optical links face growing equalization demands as symbol rates and transmission distances increase. Chromatic dispersion, bandwidth limitations, and device nonlinearities introduce intersymbol interference (ISI) that classical feed-forward equalizers (FFEs) and Volterra-type nonlinear equalizers can address, but at rapidly increasing implementation costs \cite{Savory2008,DaRos2021_SPIE,Wang2021_OptCom}. This has motivated research into machine-learning-based equalizers that can offer better performance–complexity trade-offs \cite{DaRos2021_SPIE,Wang2021_OptCom}.

Reservoir computing (RC) is attractive in this context. It consists of a fixed dynamical system (the reservoir) that maps the input waveform to a high-dimensional state. This fixed recurrent neural network is followed by a trainable linear readout layer that implements the task-specific learning \cite{Maass2002_NeuralComputation,Jaeger2001_ESN,Lukosevicius2009_CSR}. RC performance depends on the balance between memory and nonlinear processing of past inputs \cite{Dambre2012_SciRep,Inubushi2017_SciRep}—a property directly relevant to equalization. Photonic implementations are particularly compelling given their high bandwidth, low latency, and rich transient dynamics. Early demonstrations established photonic RC in delay-based and integrated platforms \cite{Vandoorne2008_OE,Appeltant2011_NatCommun,Paquot2012_SciRep,Vandoorne2014_NatCommun}, and subsequent work extended these toward increasingly capable processors \cite{Katumba2018_SciRep,Nakajima2021_CommsPhys,Shen2023_Optica,Wang2024_NatCommun}. Reservoir-based equalization has since been demonstrated in both IM/DD and coherent links \cite{Katumba2019_JLT,DaRos2020_JSTQE,Wang2021_OptCom,Sackesyn2021_OE,Masaad2024_JLT,VanAssche2025_arXiv}, and integrated photonic neural equalizers have shown similar promise \cite{Staffoli2023_PR,Staffoli2025_JLT,Wang2025_BeyondTbps}.
\begin{table*}[t]
\centering
\caption{First-order hardware accounting for the matched intra-architecture comparison. Here $N$ is the jointly trained reservoir size and $P$ is the number of programmable input weights. In the matched comparisons considered here, the $N$-node jointly trained input--readout system and the $2N$-node readout-only system use the same total number of weights. The recurrent-link count is shown as a first-order approximation for reservoir-core interconnect complexity.}
\label{tab:hardware_accounting}
\begin{tabular}{lccccc}
\hline
Architecture & Reservoir nodes & Recurrent links (approx.) & Readout weights & Input weights & Total Weights \\
\hline
Readout-only, $2N$ nodes & $2N$ & $\sim 4N$ & $2N$ complex & 0 & $2N$ \\
Joint input+readout, $N$ nodes & $N$ & $\sim 2N$ & $N$ complex & $N$ complex & $2N$ \\
\hline
\end{tabular}
\end{table*}
Despite this progress, the input transformation in most RC equalization studies is treated as fixed, random, or heuristic \cite{Appeltant2011_NatCommun,Paquot2012_SciRep,VanDerSande2017_Nanophotonics}. Yet input encoding choices matter: deterministic masks can outperform random masking \cite{Appeltant2014_SciRep}, mask temporal structure strongly affects performance \cite{Kuriki2018_OE}, timing relations between injected signals and reservoir delay influence memory capacity \cite{Stelzer2020_NeuralNetworks,Hulser2022_OME}, delayed-input strategies can improve physical RC systems \cite{Jaurigue2021_Entropy,Picco2025_CommsEng}, and input training can meaningfully improve performance compared to random input weighting in photonic computing systems \cite{BrunPNN}.

These results motivate further study in treating the input layer as a design variable, especially in the context of telecom equalization problems. This paper studies that problem in a four-port all-optical silicon photonic reservoir equalizer \cite{sackesyn2018} with integrated optical input and readout layers. Within this fixed architecture, the central question is whether the joint optimization  of input and output weights changes the node-scaling behavior strongly enough to reduce the node count required to reach a given equalization performance.

This is architecturally relevant because, in integrated photonic reservoirs, node count directly determines the number of interferometric elements, recurrent waveguide interconnects, and routing overhead. In the  comparisons considered here, an $N$-node jointly trained system uses the same total number of weights as a readout-only $2N$-node system, but uses only half the amount of components inside the reservoir layer. A comparison between the hardware needed for both scenarios is shown in table \ref{tab:hardware_accounting}.

The rest of this paper is structured as follows. In section \ref{sec:system} the hybrid simulation framework that was used to generate high-speed telecom signals and process them is explained, along with the optimisation scheme and baselines. Next, in section \ref{sec:results}, simulations of varying fiber lengths, launch power and reservoir sizes can be found along with their discussion. Finally, section \ref{sec:conc} concludes the paper.

\section{Optical Reservoir Simulations}
\label{sec:system}

This section introduces the hybrid simulation framework and the comparison protocol used to support the architectural node-scaling claim of this paper. The optical communication link is simulated in VPIphotonics \cite{VPI}, while the all-optical reservoir is simulated in the Photontorch Python library \cite{photontorch}. This separation allows transmission impairments and the photonic equalizer to be modeled using tools that are well matched to their respective physical domains.

\begin{figure*}[h!]
    \centering
    \includegraphics[width=1\textwidth]{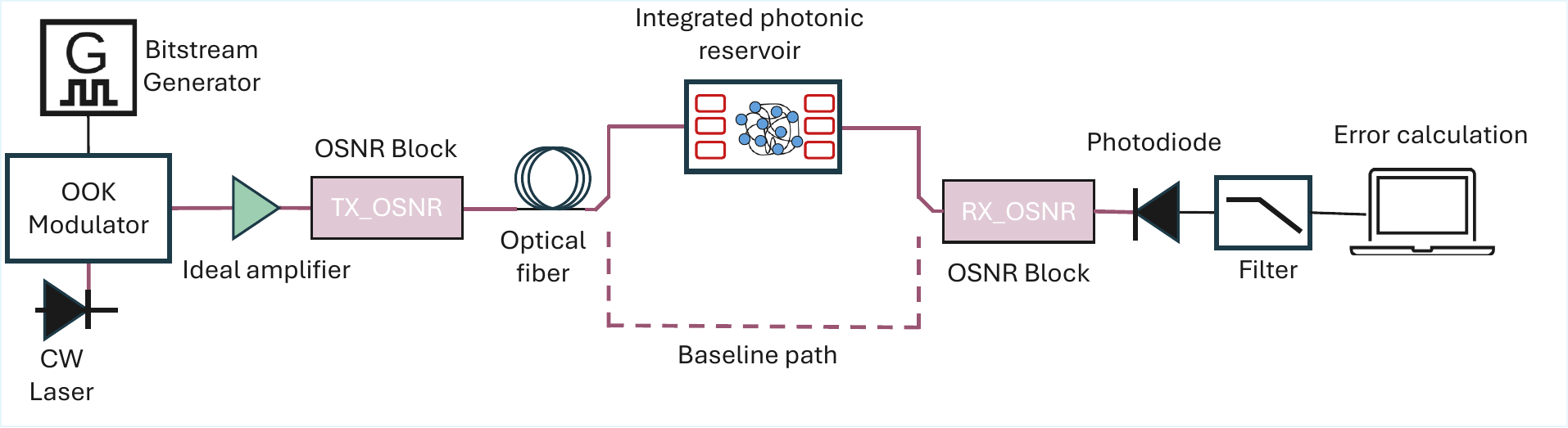}
    \caption{\textbf{Optical communications setup.} Upper and lower paths respectively show the setups with and without the photonic reservoir. The lower path is used for contextual baseline comparisons. CW: continuous wave. OOK: on-off keying. OSNR: optical signal-to-noise ratio.}
    \label{fig:setup}
\end{figure*}

\subsection{Optical Communication Model in VPIphotonics}

The distorted communication signal used as input to the reservoir equalizer is generated using VPIphotonics. The considered transmitter--receiver chain is shown schematically in Fig.~\ref{fig:setup}. A continuous-wave laser is intensity modulated using a Mach--Zehnder modulator (MZM) driven by the transmitted data sequence. The modulated optical signal then propagates through the link, where channel impairments such as chromatic dispersion, Kerr nonlinearities, and controlled noise loading are introduced.

The resulting received waveform contains the cumulative effect of transmitter modulation, optical propagation, and receiver-side noise loading. In discrete time, the detected communication signal can be written in the generic form
\begin{equation}
r_k = \sum_{i=0}^{L} h_i s_{k-i} + \eta_k + \epsilon_k,
\end{equation}
where $s_k$ denotes the transmitted symbol, $h_i$ the effective linear channel response, $L$ the channel memory, $\eta_k$ additive noise, and $\epsilon_k$ residual nonlinear distortion. The equalization task is to approximate the inverse mapping
\begin{equation}
\hat{s}_k = F(r_k,r_{k-1},\dots,r_{k-L}).
\end{equation}

In general, we use NRZ OOK transmission at \SI{28}{GBd} with 8 samples per symbol and sweep operating points over transmission distance, launch power, and receiver-side OSNR.

\subsection{Reservoir Model in Photontorch}
\label{subsec:photontorch_model}

\begin{figure*}
    \centering
    \includegraphics[width=0.9\linewidth]{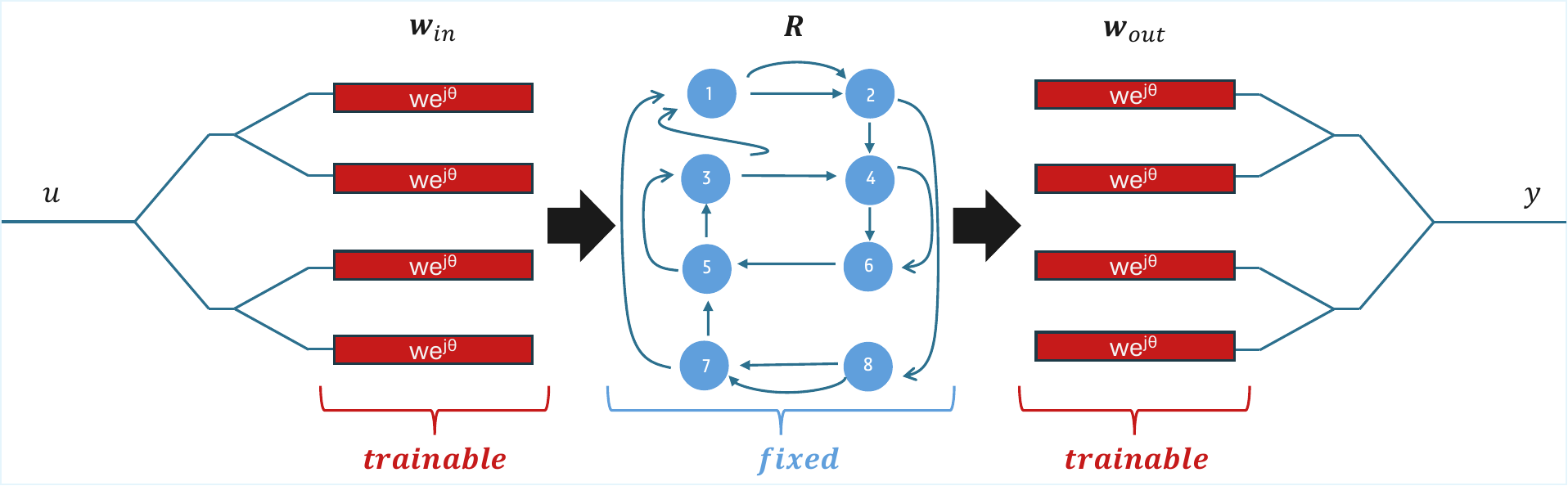}
    \caption{Schematic overview of the four-port reservoir computing architecture. An input signal $u$ gets transmitted to a trainable input layer with weights $\mathbf{w}_{in}$. This input is used in a fixed, random reservoir, here configured in the four-port architecture. Finally the states from the reservoir are extracted and recombined into an output $y$ by being transmitted through a trainable output layer with weights $w_{out}$.}
    \label{fig:4p}
\end{figure*}
The equalizer, see figure \ref{fig:4p}, takes the abstract form
\begin{equation}
\mathbf{x}(t+dt)=\mathbf{R}\mathbf{x}(t)+\mathbf{w}_{in}^{T}\mathbf{u}(t), \quad
{y}(t)=g\!\left(\mathbf{w}_{\mathrm{out}}^{T}\mathbf{x}(t)\right),
\label{eq-reservoir}
\end{equation}
where $\mathbf{x}$ is the reservoir state, $\mathbf{R}$ the fixed weight matrix of the reservoir, $\mathbf{w}_{in}$ the fixed input coupling (trainable under joint optimization), $\mathbf{u}$ the injected optical field, $\mathbf{w}_{\mathrm{out}}$ the optical readout weights, and $g(\cdot)$ the photodetection nonlinearity. While the physical reservoir computing operates in continuous time, simulations are ran in discrete time with 8 samples per symbol.

Crucially, $\mathbf{R}$ is not a trainable weight matrix: internal mixing arises physically from waveguide propagation, with each interconnect contributing attenuation, phase accumulation, and propagation delay. Fabrication-induced variations—waveguide-length errors, sidewall roughness—make each reservoir realization unique but fixed \cite{Masaad2024_JLT,VanAssche2025_arXiv}.

The inter-node waveguides within the reservoir act as delay lines set to half the target bit period, a choice validated in both simulation and experiment as effectively balancing signal mixing and temporal memory \cite{Masaad2024_JLT,VanAssche2025_arXiv}. System memory is thus determined by the physical propagation network rather than trainable recurrent weights. 

The equalizer is modeled in Photontorch as a passive spatially distributed photonic network following the four-port architecture \cite{sackesyn2018,Masaad2024_JLT,VanAssche2025_arXiv}. This architecture is a planar, power-efficient evolution of the swirl reservoir topology in which every node participates in the recurrent network through exactly two inter-node input and two inter-node output connections, eliminating the asymmetric edge splitters and associated radiation losses of the original swirl layout \cite{Dambre2021_IntegratedReservoirs,sackesyn2018}.
Each node is implemented as a $3\times3$ multimode interferometer (MMI) with two ports receiving fields from neighboring nodes, two ports routing fields onward, one external injection port, and one readout port \cite{Masaad2024_JLT}. The reservoir thus supports recurrent signal circulation, optical input injection, and state extraction within a compact planar topology.

\subsection{Programmable Optical Input and Output Layers}

The new trainable degree of freedom considered in this work is the optical input layer. Rather than injecting the distorted signal through a fixed coupling arrangement, the incoming optical field is distributed across the driven reservoir input channels using programmable complex optical weights implemented by Mach--Zehnder interferometers (MZIs).

Let $u_k\in\mathbb{C}$ denote the incoming optical field launched toward the equalizer. The optical input layer maps this field to the driven reservoir input channels according to
\begin{equation}
\mathbf{w_{in}^T}\,u(t),
\end{equation}
where $\mathbf{w}^T_{\mathrm{in}}\in\mathbb{C}^{P\times 1}$ is the trainable vector of effective input weights and $P$ is the number of input channels.

Each input weight is represented as
\begin{equation}
w_{in,i} = a_i e^{j\phi_i},
\end{equation}
with amplitude $a_i$ and phase $\phi_i$. Because the input layer is passive, the amplitude is bounded by the maximum transmission allowed by the insertion loss of the interferometer,
\begin{equation}
0 \le a_i \le A_{\max}.
\end{equation}
Hence, the trainable optical input layer is constrained to physically realizable passive complex weights of the form
\begin{equation}
\mathbf{w}_{\mathrm{in}} =
\begin{bmatrix}
a_1 e^{j\phi_1} \\
a_2 e^{j\phi_2} \\
\vdots \\
a_P e^{j\phi_P}
\end{bmatrix},
\qquad 0 \le a_i \le A_{\max}.
\end{equation}

The optical output (i.e. readout) stage combines the reservoir states coherently according to
\begin{equation}
y(t) = \mathbf{w}_{\mathrm{out}}^{T}\mathbf{x}(t),
\end{equation}
where $\mathbf{w}_{\mathrm{out}}\in\mathbb{C}^{N}$ contains the complex readout weights. The resulting optical field is then photodetected, so the final electrical output depends nonlinearly on the coherent optical recombination. The processor is therefore not an optical front-end followed by a digital linear regressor, but an all-optical computation block in which both the input projection and the readout projection are implemented in the optical domain. 

For the node-scaling experiments, we consider three input-node policies: \emph{all}, \emph{center}, and \emph{spread}. In the \emph{all} policy, all available input nodes are driven. The \emph{center} and \emph{spread} policies instead select a subset of input nodes from the full reservoir grid.

The \emph{center} policy is designed to concentrate the driven inputs near the middle of the reservoir. It first searches for a rectangular region inside the full reservoir that can contain the required number of driven nodes. Among all such candidate rectangles, it prefers the most compact one; if several are equally compact, it chooses the one whose shape best matches the overall reservoir aspect ratio. This rectangle is then placed at the center of the reservoir grid, and the driven nodes are selected from the middle outward so that the chosen inputs remain as concentrated as possible around the geometric center.

The \emph{spread} policy follows the opposite principle. It is designed to distribute the driven inputs as broadly as possible over the reservoir grid. Starting from a node near the outer part of the reservoir, additional driven nodes are added one by one using a greedy farthest-point rule based on Euclidean distance on the two-dimensional grid. At each step, the next node is chosen to maximize its distance from the nodes already selected, yielding a set of driven inputs that is distributed across the full reservoir footprint.
\subsection{Training Objective, Node-Scaling Comparison, and Fairness}
\label{subsec:training_protocol}

The trainable parameters are optimized using the Broyden--Fletcher--Goldfarb--Shanno algorithm (BFGS).

For each set of weights, the VPI-generated waveform is propagated through the Photontorch reservoir. A warm-up interval is discarded to suppress transient effects. The remaining data are split into 8192 training symbols, a validation set of 32768 sybmols, and a test set consisting of 10 blocks of 32768 symbols all generated with a different random seed. Receiver-side conditioning before evaluation consists of low-pass filtering at \SI{28}{GHz} followed by normalization.

The training objective is the mean-squared error (MSE) evaluated at the symbol decision points. To find the optimal decision point, a discrete output-target latency sweep is performed to align the equalizer output to the target symbol sequence. This sweep ranges from 0 to 160 samples, corresponding to 20 symbol intervals at 8 samples per symbol. The selected latency is an alignment parameter of the training and evaluation pipeline; it does not modify the physical delay structure of the reservoir itself. While we use the MSE for training, for testing and validation, the BER and statistical BER \cite{statber} are used as  metrics.

The central comparison of this paper is between matched variants of reservoirs with a different size:
\begin{enumerate}
    \item an $N$-node reservoir with jointly optimized input and output (readout) weights;
    \item a readout-only reservoir with $2N$ nodes.
\end{enumerate}
The purpose is to determine whether the performance gain obtained by joint optimization is large enough to offset a factor-of-two reduction in reservoir size.

All reservoir variants use the same communication waveforms, train/test split, latency sweep, receiver-side filtering and normalization, and the same evaluation metrics. 

To evaluate whether these findings remain consistent over different random reservoir initialisations, the simulations for each operating point are repeated over 10 random reservoir phase realizations. In readout-only experiments, the fixed input phases are additionally randomized using a separate random seed to avoid an artificially symmetric excitation pattern. 

For each latency, both the readout-only and joint input--readout optimizations use the same BFGS algorithm, latency-selection protocol,  hyperparameter sweep over initialization radius, and the same maximum training budget.  In practice, both scenarios were observed to reach comparably good local optima under this common protocol, although the joint optimization typically required more function evaluations. It should be noted that both in the case of the joint training or the readout-only training, the optimizer always converged to a solution. The proposed method should therefore not be interpreted as reducing training complexity. Rather, it explicitly trades greater offline optimization effort for a reduction in the reservoir size required to reach a given equalization regime.

For comparisons based on the model size, we count the number of real-valued trainable optical degrees of freedom. Each complex optical weight therefore contributes two real degrees of freedom, corresponding to its effective amplitude and phase parameters. Unless stated otherwise, degree-of-freedom counts refer only to trainable optical weights and not to fixed architectural choices such as input-node policy.

For context and benchmarking, we also implemented a linear FFE and a second-order nonlinear Volterra equalizer on the received waveform. In both cases, taps were spaced by half a bit period, matching the temporal sampling granularity used for the reservoir computer. The number of FFE taps was chosen such that the number of real-valued trainable coefficients matches the number of real-valued trainable degrees of freedom of the optical network. The second-order Volterra baseline was instead chosen to match the same memory span as the FFE and therefore has a quadratically larger number of trainable degrees of freedom.

\section{Results}
\label{sec:results}

\begin{figure*}[h!]
    \centering
    \includegraphics[width=1\textwidth]{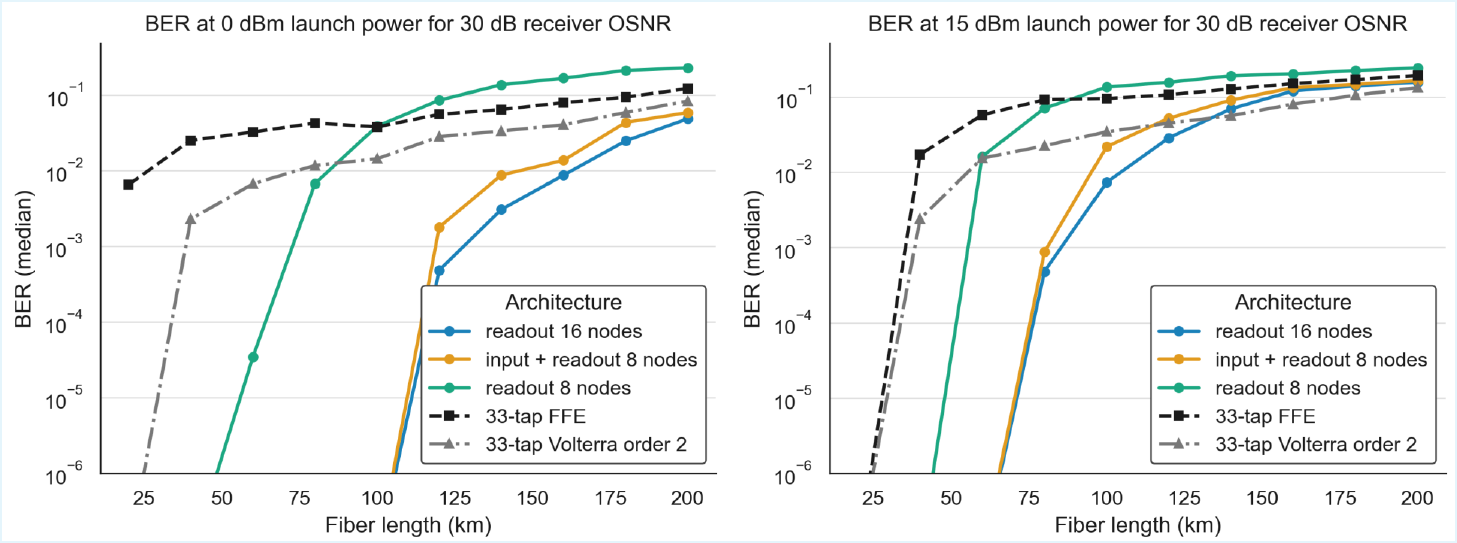}
    \caption{Median BER versus fiber length at (a) 0 and (b) 15~dBm launch power and 30~dB receiver OSNR for an 8-node jointly optimized reservoir, an 8-node readout-only reservoir, and a 16-node readout-only reservoir. The jointly optimized 8-node variant consistently improves on the 8-node readout-only variant and, over a useful short-to-intermediate-length regime, approaches the BER range of the 16-node readout-only variant.}
    \label{fig:length_sweep_15dbm}
\end{figure*}
\subsection{Fiber-length sweeps at fixed receiver OSNR}
\label{subsec:length_sweeps}

We first compare three systems as a function of transmission distance at a fixed receiver OSNR of 30~dB: an 8-node reservoir with joint input-readout optimization, an 8-node readout-only variant, and a 16-node readout-only variant.

Figure~\ref{fig:length_sweep_15dbm}(a) shows the median BER as a function of fiber length for a launch power of 10~dBm. Several trends are clear. First, joint optimization provides a substantial improvement over the 8-node readout-only variant across the full sweep. Across all lengths, the median BER of the jointly trained 8 node RC is roughly one to two orders of magnitude below that of readout trained 8 node RC. It is also clear that a gap smaller than an order of magnitude persists between the jointly trained 8 node RC and the readout trained 16 node RC. Thus, over the short-to-intermediate-length regime, jointly optimizing the optical input and readout allows an 8-node reservoir to reach the same BER order as a 16-node readout-only variant with half the amount of reservoir hardware.

Beyond approximately 180~km, all three systems deteriorate sharply and the gap between them narrows. In this regime, the underlying channel impairment becomes too strong for the smaller jointly optimized system and the 16-node readout-only system, as both move into a comparatively high-error regime. This indicates that joint optical-layer optimization can expose more useful reservoir features, but does not remove the operating-window limitations of the architecture and the memory requirements of the task. Compared to the baseline FFE and Volterra filter, it can be seen that the reservoirs with an equal number of degrees of freedom consistently outperform these.

The corresponding sweep at 15~dBm launch power is shown in Fig.~\ref{fig:length_sweep_15dbm}(b). The same qualitative ranking is observed, where the jointly trained 8 node RC consistently outperforms the 8-node readout-only variant and follows the performance of the readout-only 16 node RC. It can also be seen that for this more nonlinear and difficult task, at high lengths the second-order Volterra filter with 33 taps, benefiting from its higher nonlinear computational power, does perform at the same order of magnitude as the RC's starting at 140 km, where performance in general collapses to the order of $10^{-1}$. Taken together, Fig.~\ref{fig:length_sweep_15dbm}(a) and Fig.~\ref{fig:length_sweep_15dbm}(b) show that the benefit of trainable optical encoding is robust within the explored operating window.

\subsection{Node scaling at a challenging operating point}
\label{subsec:node_scaling}

To study the node-scaling question in more detail, we next fix a challenging operating point at 200~km fiber length, 15~dBm launch power, and 30~dB receiver OSNR, and examine how performance evolves with reservoir size and input-node policy. The considered policies are the earlier explained all, spread and center policies both for joint training of input and readout layer, and readout-only training.
\begin{figure*}[h!]
    \centering
    \includegraphics[width=0.6\textwidth]{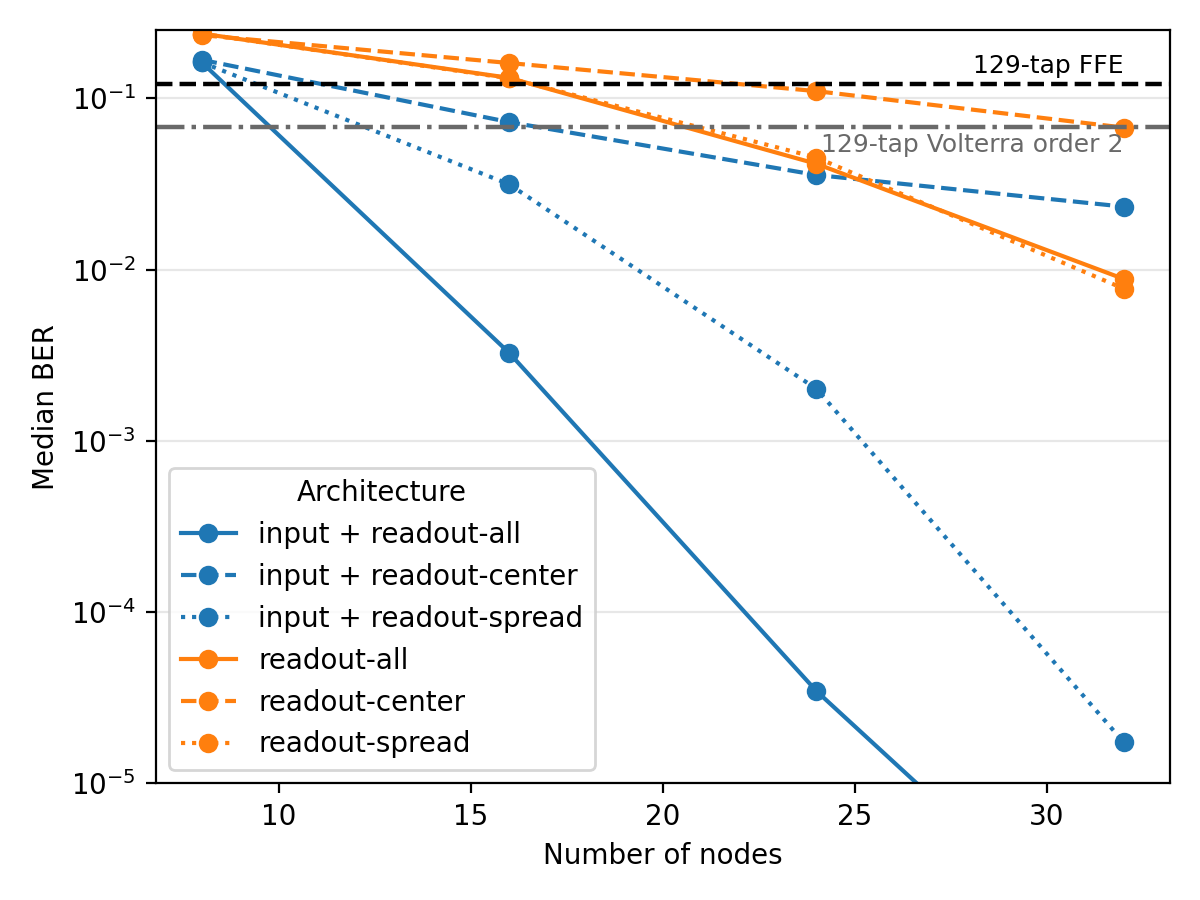}
    \caption{Median BER versus reservoir size. The jointly optimized configurations, particularly IR-all, improve much more rapidly with node count than the readout-only variants, indicating that trainable optical encoding increases the usefulness of additional reservoir nodes in this operating regime.}
    \label{fig:node_scaling_median}
\end{figure*}

\begin{table*}[t]
\centering
\caption{Representative BER statistics for the 200~km node-scaling study at 15~dBm launch power and 30~dB receiver OSNR. Each entry is the median [Q1, Q3] over 10 paired realizations. For each node count, the best readout-only median among all policies is shown.}
\label{tab:nodesweep_stats}
\begin{tabular}{lcc}
\hline
Nodes & joint input and readout training-all & Best readout-only variant at same node count \\
\hline
8  & $1.62\times10^{-1}$ [$1.58\times10^{-1}$, $1.78\times10^{-1}$] & $2.34\times10^{-1}$ [$2.15\times10^{-1}$, $2.42\times10^{-1}$] (Readout-only center) \\
16 & $3.27\times10^{-3}$ [$2.86\times10^{-3}$, $4.82\times10^{-3}$] & $1.31\times10^{-1}$ [$1.22\times10^{-1}$, $1.40\times10^{-1}$] (Readout-only spread) \\
24 & $3.54\times10^{-5}$ [$0$, $8.72\times10^{-5}$] & $4.15\times10^{-2}$ [$3.74\times10^{-2}$, $5.26\times10^{-2}$] (Readout-only all) \\
32 & $0$ [$0$, $0$] & $7.73\times10^{-3}$ [$4.52\times10^{-3}$, $9.41\times10^{-3}$] (Readout-only spread) \\
\hline
\end{tabular}
\end{table*}
Figure~\ref{fig:node_scaling_median} shows the node-evolution of the median BER over the 10 paired seed realizations. The most important observation is that the jointly optimized systems improve much more rapidly with node count than the readout-only variants. In particular, the jointly trained all configuration already reaches a much lower error regime at 16 nodes than any of the 16-node readout-only variants, and by 24 nodes it is separated from all readout-only systems by several orders of magnitude in median BER. For all readout-only variants, performance improves only gradually with increasing reservoir size. Once the optical input layer is trained jointly with the readout, the same increase in node count becomes much more effective. The jointly trained systems make better use of each additional node, indicating that the learned optical excitation exposes a more useful subset of the reservoir's delayed interference features.

Table~\ref{tab:nodesweep_stats} provides a compact numerical summary of the strongest jointly optimized policy and the strongest readout-only policy at each node count, including their spreads. 

\begin{figure*}[h!]
    \centering
    \includegraphics[width=0.6\textwidth]{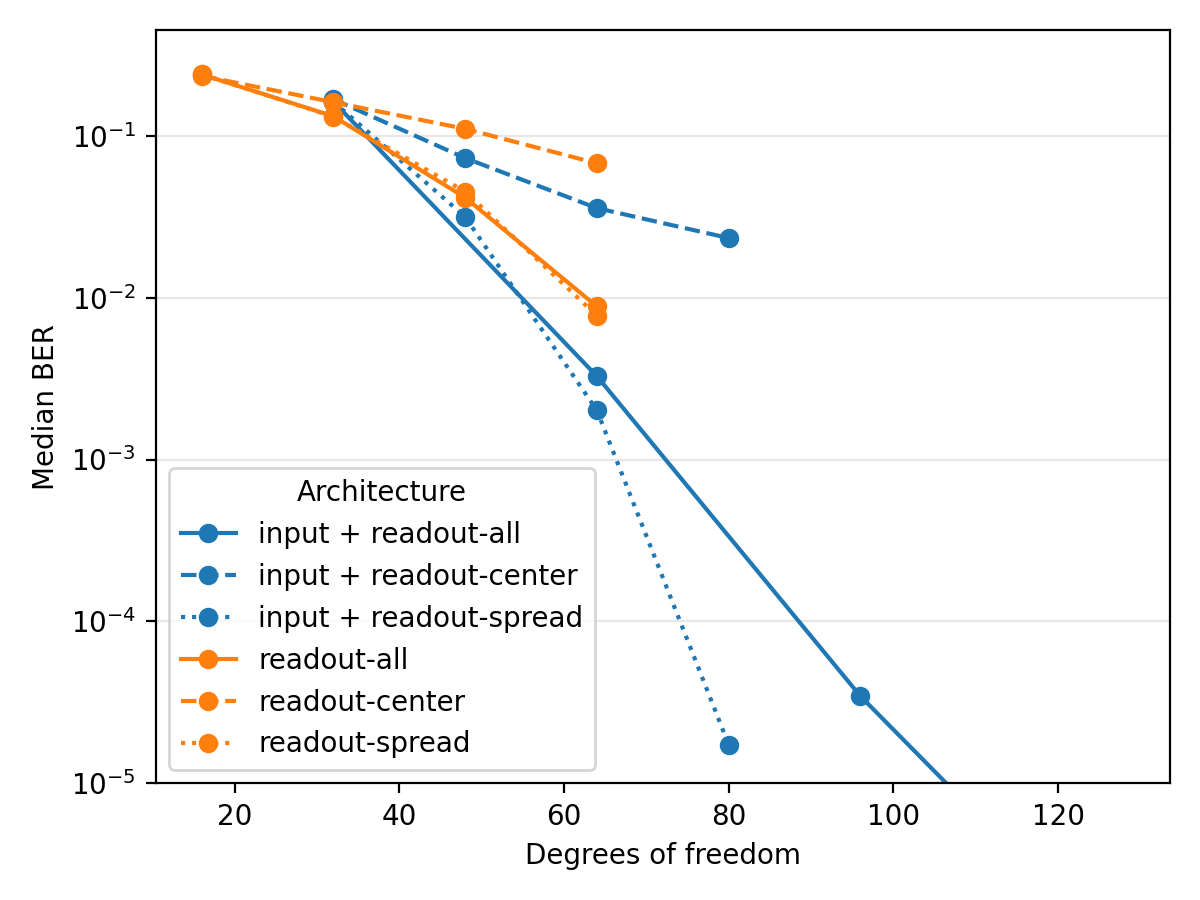}
    \caption{Median BER versus the number of real-valued trainable degrees of freedom for the same operating point as in Fig.~\ref{fig:node_scaling_median}. Each complex optical weight is counted as two real degrees of freedom. Jointly optimized input--readout architectures achieve substantially lower BER than readout-only variants at comparable trainable degrees of freedom, indicating that the observed gain is not merely a consequence of parameter count but of how the trainable optical resources are allocated within the architecture.}
    \label{fig:node_scaling_dof}
\end{figure*}

The same conclusion is supported when performance is plotted against trainable real-valued degrees of freedom rather than node count. Figure~\ref{fig:node_scaling_dof} shows that jointly optimized input-readout architectures achieve similar or lower BER than readout-only variants at comparable numbers of trainable optical degrees of freedom. This again indicates that the observed performance is not merely a consequence of parameter count, but of how the trainable optical resources are allocated within the architecture.\\

\subsection{Effect of input-node policy}
\label{subsec:input_policy_results}

The node-scaling results also show that the choice of input-node policy matters. Among the jointly optimized systems, the 'all' policy consistently provides the strongest scaling trend, while the 'spread' policy also performs well at larger node counts. By contrast, the 'center' policy improves more slowly and remains clearly above the best jointly optimized configuration. This indicates that the trainable optical front-end benefits from access to a broader set of injection channels: the input layer is most effective when it can redistribute optical amplitude and phase over multiple excitation paths and thereby reshape the delayed quadratic feature map of the reservoir.

For the readout-only variants, the differences between input-node policies are smaller and remain within a comparatively poor error regime. This again supports the interpretation that the main gain does not arise from input placement alone, but from the ability to jointly optimize how the distorted waveform is injected into the reservoir and how the resulting states are optically recombined at the output.

\section{Acknowledgements}
This work was supported by the European Commission in the Horizon 2020 programme under the project Nebula, in the Horizon Europe programme under the projects Prometheus, POSTDIGITAL+, Neho, Neuropuls and NEHIL, and by the Belgian FWO and EOS programmes.

\section{Conclusion}
\label{sec:conc}
Scaling computation on photonic integrated circuits is limited by higher hardware costs and increased complexity. We investigated whether a trainable optical input layer can change the node-scaling behavior of a four-port photonic reservoir sufficiently to reduce the photonic hardware required to reach a given equalization regime. Using a hybrid VPIphotonics--Photontorch framework, jointly optimized input--readout systems were compared against matched readout-only variants under identical simulation and evaluation conditions.
The central result is that joint optimization provides a clear node-scaling advantage within this architecture. The jointly optimized 8-node reservoir consistently outperformed its readout-only counterpart and, over a useful short-to-intermediate fiber-length regime, approached the BER of a 16-node readout-only system. At 200~km, the advantage sharpened further: jointly optimized systems scaled substantially better with node count, with the strongest configurations well beyond the range reached by readout-only training.
Interpretively, the four-port reservoir acts as a fixed delayed linear optical feature generator, with quadratic interactions arising from coherent recombination and photodetection. The trainable input layer reshapes how the fixed interference structure of the reservoir is excited and presented to the detector, giving the readout access to a more useful feature map for inverse-channel recovery. The node-scaling benefit therefore reflects a more effective use of existing reservoir dynamics, not an additional memory capacity.
For the matched comparisons considered here, an $N$-node jointly optimized architecture uses the same number of weights as a readout-only $2N$-node architecture while requiring fewer reservoir nodes and recurrent interconnects. This implies a reduction of photonic hardware by a factor of 2 in the reservoir, not a mere redistribution of trainable weights. The benefit does carry one practical cost however: greater offline optimization effort in exchange for a smaller inference-time reservoir core.

Experimental validation, explicit area-control trade-off analysis, and extension to other modulation formats and reservoir architectures are natural directions for future work.

\printbibliography

@article{Savory2008,
author = {Seb J. Savory},
journal = {Opt. Express},
number = {2},
pages = {804--817},
publisher = {Optica Publishing Group},
title = {Digital filters for coherent optical receivers},
volume = {16},
month = {Jan},
year = {2008},
url = {https://opg.optica.org/oe/abstract.cfm?URI=oe-16-2-804},
doi = {10.1364/OE.16.000804},
}

@inproceedings{DaRos2021_SPIE,
author = {F. Da Ros and S. M. Ranzini and R. Dischler and A. Cem and V. Aref and H. B{\"u}low and D. Zibar},
title = {{Machine-learning-based equalization for short-reach transmission: neural networks and reservoir computing}},
volume = {11712},
booktitle = {Metro and Data Center Optical Networks and Short-Reach Links IV},
editor = {Atul K. Srivastava and Madeleine Glick and Youichi Akasaka},
organization = {International Society for Optics and Photonics},
publisher = {SPIE},
pages = {1171205},
year = {2021},
doi = {10.1117/12.2583011},
URL = {https://doi.org/10.1117/12.2583011}
}

@article{Wang2021_OptCom,
title = {Signal recovery based on optoelectronic reservoir computing for high speed optical fiber communication system},
journal = {Optics Communications},
volume = {495},
pages = {127082},
year = {2021},
issn = {0030-4018},
doi = {https://doi.org/10.1016/j.optcom.2021.127082},
url = {https://www.sciencedirect.com/science/article/pii/S003040182100331X},
author = {Shuai Wang and Nian Fang and Lutang Wang}
}

@ARTICLE{Maass2002_NeuralComputation,
  author={Maass, Wolfgang and Natschläger, Thomas and Markram, Henry},
  journal={Neural Computation}, 
  title={Real-Time Computing Without Stable States: A New Framework for Neural Computation Based on Perturbations}, 
  year={2002},
  volume={14},
  number={11},
  pages={2531-2560},
  doi={10.1162/089976602760407955}}

@inproceedings{Jaeger2001_ESN,
  title={The “echo state” approach to analysing and training recurrent neural networks-with an erratum note},
  author={Jaeger, Herbert},
  journal={Bonn, Germany: German national research center for information technology gmd technical report},
  volume={148},
  number={34},
  pages={13},
  year={2001},
  publisher={Bonn}
}

@article{Lukosevicius2009_CSR,
title = {Reservoir computing approaches to recurrent neural network training},
journal = {Computer Science Review},
volume = {3},
number = {3},
pages = {127-149},
year = {2009},
issn = {1574-0137},
doi = {https://doi.org/10.1016/j.cosrev.2009.03.005},
url = {https://www.sciencedirect.com/science/article/pii/S1574013709000173},
author = {Mantas Lukoševičius and Herbert Jaeger}
}

@article{Dambre2012_SciRep,
author = {Dambre, J. and Verstraeten, David and Schrauwen, Benjamin and Massar, Serge},
year = {2012},
month = {07},
pages = {514},
title = {Information Processing Capacity of Dynamical Systems},
volume = {2},
journal = {Scientific reports},
doi = {10.1038/srep00514}
}

@article{Inubushi2017_SciRep,
author = {Inubushi, Masanobu and Yoshimura, Kazuyuki},
year = {2017},
month = {12},
pages = {},
title = {Reservoir Computing Beyond Memory-Nonlinearity Trade-off},
volume = {7},
journal = {Scientific Reports},
doi = {10.1038/s41598-017-10257-6}
}

@article{Vandoorne2008_OE,
author = {Kristof Vandoorne and Wouter Dierckx and Benjamin Schrauwen and David Verstraeten and Roel Baets and Peter Bienstman and Jan Van Campenhout},
journal = {Opt. Express},
number = {15},
pages = {11182--11192},
publisher = {Optica Publishing Group},
title = {Toward optical signal processing using Photonic Reservoir Computing},
volume = {16},
month = {Jul},
year = {2008},
url = {https://opg.optica.org/oe/abstract.cfm?URI=oe-16-15-11182},
doi = {10.1364/OE.16.011182},
}

@article{Appeltant2011_NatCommun,
  title={Information processing using a single dynamical node as complex system},
  author={Lennert Appeltant and Miguel C. Soriano and Guy Van der Sande and Jan Danckaert and Serge Massar and Joni Dambre and Benjamin Schrauwen and Claudio R. Mirasso and Ingo Fischer},
  journal={Nature Communications},
  year={2011},
  volume={2},
  url={https://api.semanticscholar.org/CorpusID:7571855}
}

@article{Paquot2012_SciRep,
  author  = {Yvan Paquot and Fran\c{c}ois Duport and Anteo Smerieri and Joni Dambre and Benjamin Schrauwen and Marc Haelterman and Serge Massar},
  title   = {Optoelectronic Reservoir Computing},
  journal = {Scientific Reports},
  year    = {2012},
  volume  = {2},
  pages   = {287},
  doi     = {10.1038/srep00287},
  url     = {https://www.nature.com/articles/srep00287}
}

@article{Vandoorne2014_NatCommun,
  title={Experimental demonstration of reservoir computing on a silicon photonics chip},
  author={Kristof Vandoorne and Pauline Mechet and Thomas Van Vaerenbergh and Martin Fiers and Geert Morthier and David Verstraeten and Benjamin Schrauwen and Joni Dambre and Peter Bienstman},
  journal={Nature Communications},
  year={2014},
  volume={5},
  url={https://api.semanticscholar.org/CorpusID:205324293}
}

@article{VanDerSande2017_Nanophotonics,
url = {https://doi.org/10.1515/nanoph-2016-0132},
title = {Advances in photonic reservoir computing},
title = {},
author = {Guy Van der Sande and Daniel Brunner and Miguel C. Soriano},
pages = {561--576},
volume = {6},
number = {3},
journal = {Nanophotonics},
doi = {doi:10.1515/nanoph-2016-0132},
year = {2017},
lastchecked = {2026-03-10}
}

@article{Katumba2018_SciRep,
author = {Katumba, Andrew and Heyvaert, Jelle and Schneider, Bendix and Uvin, Sarah and Dambre, J. and Bienstman, Peter},
year = {2018},
month = {02},
pages = {},
title = {Low-Loss Photonic Reservoir Computing with Multimode Photonic Integrated Circuits},
volume = {8},
journal = {Scientific Reports},
doi = {10.1038/s41598-018-21011-x}
}

@article{Nakajima2021_CommsPhys,
  title={Scalable reservoir computing on coherent linear photonic processor},
  author={Mitsumasa Nakajima and Kenji Tanaka and Toshikazu Hashimoto},
  journal={Communications Physics},
  year={2021},
  volume={4},
  url={https://api.semanticscholar.org/CorpusID:231876989}
}

@article{Sackesyn2021_OE,
author = {Stijn Sackesyn and Chonghuai Ma and Joni Dambre and Peter Bienstman},
journal = {Opt. Express},
number = {20},
pages = {30991--30997},
publisher = {Optica Publishing Group},
title = {Experimental realization of integrated photonic reservoir computing for nonlinear fiber distortion compensation},
volume = {29},
month = {Sep},
year = {2021},
url = {https://opg.optica.org/oe/abstract.cfm?URI=oe-29-20-30991},
doi = {10.1364/OE.435013},
}

@article{Shen2023_Optica,
author = {Yi-Wei Shen and Rui-Qian Li and Guan-Ting Liu and Jingyi Yu and Xuming He and Lilin Yi and Cheng Wang},
journal = {Optica},
number = {12},
pages = {1745--1751},
publisher = {Optica Publishing Group},
title = {Deep photonic reservoir computing recurrent network},
volume = {10},
month = {Dec},
year = {2023},
url = {https://opg.optica.org/optica/abstract.cfm?URI=optica-10-12-1745},
doi = {10.1364/OPTICA.506635},
}

@article{Wang2024_NatCommun,
  title={Ultrafast silicon photonic reservoir computing engine delivering over 200 TOPS},
  author={Dongliang Wang and Yikun Nie and Gaolei Hu and Hon Ki Tsang and Chaoran Huang},
  journal={Nature Communications},
  year={2024},
  volume={15},
  url={https://api.semanticscholar.org/CorpusID:275142268}
}

@ARTICLE{Katumba2019_JLT,
  author={Katumba, Andrew and Yin, Xin and Dambre, Joni and Bienstman, Peter},
  journal={Journal of Lightwave Technology}, 
  title={A Neuromorphic Silicon Photonics Nonlinear Equalizer For Optical Communications With Intensity Modulation and Direct Detection}, 
  year={2019},
  volume={37},
  number={10},
  pages={2232-2239},
  doi={10.1109/JLT.2019.2900568}}

@ARTICLE{DaRos2020_JSTQE,
  author={Da Ros, Francesco and Ranzini, Stenio M. and Bülow, Henning and Zibar, Darko},
  journal={IEEE Journal of Selected Topics in Quantum Electronics}, 
  title={Reservoir-Computing Based Equalization With Optical Pre-Processing for Short-Reach Optical Transmission}, 
  year={2020},
  volume={26},
  number={5},
  pages={1-12},
  doi={10.1109/JSTQE.2020.2975607}}

@ARTICLE{Masaad2024_JLT,
  author={Masaad, Sarah and Sackesyn, Stijn and Sygletos, Stylianos and Bienstman, Peter},
  journal={Journal of Lightwave Technology}, 
  title={Experimental Demonstration of 4-Port Photonic Reservoir Computing for Equalization of 4 and 16 QAM Signals}, 
  year={2024},
  volume={42},
  number={24},
  pages={8555-8563},
  doi={10.1109/JLT.2024.3444480}}

@misc{VanAssche2025_arXiv,
      title={Real-time all-optical signal equalisation with silicon photonic recurrent neural networks}, 
      author={Ruben Van Assche and Sarah Masaad and Emmanuel Gooskens and Stijn Sackesyn and Joris Van Kerrebrouck and Xin Yin and Peter Bienstman},
      year={2025},
      eprint={2503.19911},
      archivePrefix={arXiv},
      primaryClass={physics.optics},
      url={https://arxiv.org/abs/2503.19911}, 
}

@article{Appeltant2014_SciRep,
title = "Constructing optimized binary masks for reservoir computing with delay systems",
author = "Lennert Appeltant and {Van Der Sande}, Guy and Jan Danckaert and Ingo Fischer",
year = "2014",
month = jan,
day = "10",
doi = "10.1038/srep03629",
language = "English",
volume = "4",
journal = "Scientific Reports - Nature",
issn = "2045-2322",
publisher = "Nature Publishing Group",

}

@article{Kuriki2018_OE,
author = {Yoma Kuriki and Joma Nakayama and Kosuke Takano and Atsushi Uchida},
journal = {Opt. Express},
number = {5},
pages = {5777--5788},
publisher = {Optica Publishing Group},
title = {Impact of input mask signals on delay-based photonic reservoir computing with semiconductor lasers},
volume = {26},
month = {Mar},
year = {2018},
url = {https://opg.optica.org/oe/abstract.cfm?URI=oe-26-5-5777},
doi = {10.1364/OE.26.005777},
}

@article{Stelzer2020_NeuralNetworks,
title = {Performance boost of time-delay reservoir computing by non-resonant clock cycle},
journal = {Neural Networks},
volume = {124},
pages = {158-169},
year = {2020},
issn = {0893-6080},
doi = {https://doi.org/10.1016/j.neunet.2020.01.010},
url = {https://www.sciencedirect.com/science/article/pii/S0893608020300125},
author = {Florian Stelzer and André Röhm and Kathy Lüdge and Serhiy Yanchuk}
}

@article{Hulser2022_OME,
author = {Tobias H\"{u}lser and Felix K\"{o}ster and Lina Jaurigue and Kathy L\"{u}dge},
journal = {Opt. Mater. Express},
number = {3},
pages = {1214--1231},
publisher = {Optica Publishing Group},
title = {Role of delay-times in delay-based photonic reservoir computing, Invited},
volume = {12},
month = {Mar},
year = {2022},
url = {https://opg.optica.org/ome/abstract.cfm?URI=ome-12-3-1214},
doi = {10.1364/OME.451016},
}

@Article{Jaurigue2021_Entropy,
AUTHOR = {Jaurigue, Lina and Robertson, Elizabeth and Wolters, Janik and Lüdge, Kathy},
TITLE = {Reservoir Computing with Delayed Input for Fast and Easy Optimisation},
JOURNAL = {Entropy},
VOLUME = {23},
YEAR = {2021},
NUMBER = {12},
ARTICLE-NUMBER = {1560},
URL = {https://www.mdpi.com/1099-4300/23/12/1560},
PubMedID = {34945866},
ISSN = {1099-4300},
DOI = {10.3390/e23121560}
}

@article{Picco2025_CommsEng,
  author  = {Enrico Picco and Lina C. Jaurigue and Kathy L\"udge and Serge Massar},
  title   = {Efficient optimisation of physical reservoir computers using only a delayed input},
  journal = {Communications Engineering},
  year    = {2025},
  volume  = {4},
  number  = {1},
  pages   = {3},
  doi     = {10.1038/s44172-025-00340-6},
  url     = {https://www.nature.com/articles/s44172-025-00340-6}
}

@article{Staffoli2023_PR,
author = {Emiliano Staffoli and Mattia Mancinelli and Paolo Bettotti and Lorenzo Pavesi},
journal = {Photon. Res.},
number = {5},
pages = {878--886},
publisher = {Optica Publishing Group},
title = {Equalization of a 10\&\#x2009;\&\#x2009;Gbps IMDD signal by a small silicon photonics time delayed neural network},
volume = {11},
month = {May},
year = {2023},
url = {https://opg.optica.org/prj/abstract.cfm?URI=prj-11-5-878},
doi = {10.1364/PRJ.483356},
}

@article{Staffoli2025_JLT,
author = {Emiliano Staffoli and Gianpietro Maddinelli and Lorenzo Pavesi},
journal = {J. Lightwave Technol.},
number = {2},
pages = {557--571},
publisher = {Optica Publishing Group},
title = {A Silicon Photonic Neural Network for Chromatic Dispersion Compensation in 20 Gbps PAM4 Signal at 125 km and its Scalability up to 100 Gbps},
volume = {43},
month = {Jan},
year = {2025},
url = {https://opg.optica.org/jlt/abstract.cfm?URI=jlt-43-2-557},
}

@misc{Wang2025_BeyondTbps,
      title={Beyond Terabit/s Integrated Neuromorphic Photonic Processor for DSP-Free Optical Interconnects}, 
      author={Benshan Wang and Qiarong Xiao and Tengji Xu and Li Fan and Shaojie Liu and Jianji Dong and Junwen Zhang and Chaoran Huang},
      year={2025},
      eprint={2504.15044},
      archivePrefix={arXiv},
      primaryClass={physics.optics},
      url={https://arxiv.org/abs/2504.15044}, 
}

@inproceedings{sackesyn2018,
  author       = {{Sackesyn, Stijn and Ma, Chonghuai and Dambre, Joni and Bienstman, Peter}},
  booktitle    = {{Cognitive Computing 2018 - Merging Concepts with Hardware}},
  language     = {{und}},
  location     = {{Hannover, Germany}},
  pages        = {{1--2}},
  title        = {{An enhanced architecture for silicon photonic reservoir computing}},
  year         = {{2018}},
}

@incollection{Dambre2021_IntegratedReservoirs,
  author       = {{Dambre, Joni and Katumba, Andrew and Ma, Chonghuai and Sackesyn, Stijn and Laporte, Floris and Freiberger, Matthias and Bienstman, Peter}},
  booktitle    = {{Reservoir computing : theory, physical implementations, and applications}},
  editor       = {{Nakajima, Kohei and Fischer, Ingo}},
  isbn         = {{9789811316869}},
  issn         = {{1619-7127}},
  language     = {{eng}},
  pages        = {{397--419}},
  publisher    = {{Springer}},
  series       = {{Natural Computing Series}},
  title        = {{Computing with integrated photonic reservoirs}},
  url          = {{http://doi.org/10.1007/978-981-13-1687-6_17}},
  year         = {{2021}},
}

@software{VPI,
  author = {{VPIphotonics}},
  title = {VPI Design Suite},
  url = {https://www.vpiphotonics.com/Tools/DesignSuite/},
  version = {11.5, Accessed Mar. 03, 2026},
}

@article{photontorch,
  title={Highly parallel simulation and optimization of photonic circuits in time and frequency domain based on the deep-learning framework pytorch},
  author={Laporte, Floris and Dambre, Joni and Bienstman, Peter},
  journal={Scientific reports},
  volume={9},
  number={1},
  pages={5918},
  year={2019},
  publisher={Nature Publishing Group UK London}
}

@misc{BrunPNN,
      title={Model-free front-to-end training of a large high performance laser neural network}, 
      author={Anas Skalli and Satoshi Sunada and Mirko Goldmann and Marcin Gebski and Stephan Reitzenstein and James A. Lott and Tomasz Czyszanowski and Daniel Brunner},
      year={2025},
      eprint={2503.16943},
      archivePrefix={arXiv},
      primaryClass={cs.LG},
      url={https://arxiv.org/abs/2503.16943}, 
}

@ARTICLE{statber,
  author={Bergano, N.S. and Kerfoot, F.W. and Davidsion, C.R.},
  journal={IEEE Photonics Technology Letters}, 
  title={Margin measurements in optical amplifier system}, 
  year={1993},
  volume={5},
  number={3},
  pages={304-306},
  doi={10.1109/68.205619}}
\end{document}